\documentclass[pdflatex,sn-mathphys-num]{sn-jnl}% Math and Physical Sciences Numbered Reference Style 
\usepackage{graphicx}%
\usepackage{multirow}%
\usepackage{amsmath,amssymb,amsfonts}%
\usepackage{amsthm}%
\usepackage{mathrsfs}%
\usepackage[title]{appendix}%
\usepackage{xcolor}%
\usepackage{textcomp}%
\usepackage{hyperref}
\usepackage{manyfoot}%
\usepackage{booktabs}%
\usepackage{algorithm}
\usepackage{algpseudocode}
\usepackage{algorithm}%
\usepackage{algorithmicx}%
\usepackage{algpseudocode}%
\usepackage{listings}%
\usepackage{braket}
\theoremstyle{thmstyleone}%
\theoremstyle{thmstyletwo}%

\theoremstyle{thmstylethree}%

\begin{document}

\title[]{Simulating and Learning Quantum Evolution: A CTQW-ML Framework }

%%=============================================================%%
%% GivenName	-> \fnm{Joergen W.}
%% Particle	-> \spfx{van der} -> surname prefix
%% FamilyName	-> \sur{Ploeg}
%% Suffix	-> \sfx{IV}
%% \author*[1,2]{\fnm{Joergen W.} \spfx{van der} \sur{Ploeg} 
%%  \sfx{IV}}\email{iauthor@gmail.com}
%%=============================================================%%

\author*[1]{\fnm{Rachana} \sur{Soni}}\email{rachanasoni007@gmail.com}

% \author[2]{\fnm{Neelam} \sur{Choudhary}}\email{neelam.choudhary@bennett.edu.in}
% % \equalcont{These authors contributed equally to this work.}

\author[2]{\fnm{Navneet} \sur{Pratap Singh}}%\email{navneet.singh@bennett.edu.in}
% \equalcont{These authors contributed equally to this work.}

\affil*{\orgdiv{School of Computer Science Engineering and Technology}, \orgname{Bennett University}, \orgaddress{\street{TechZone 2}, \city{Greater Noida}, \postcode{201310}, \state{U.P.}, \country{India}}}

\affil[2]{\orgdiv{School of Artificial Intelligence}, \orgname{Bennett University}, \orgaddress{\street{TechZone 2}, \city{Greater Noida}, \postcode{201310}, \state{U.P.}, \country{India}}}

%%==================================%%
%% Sample for unstructured abstract %%
%%==================================%%

\abstract{We present an approach to simulate the  Schr\"odinger equation through continuous time quantum walks. The CTQW-based simulation applies unitary evolution driven by a quantum walk  to generate probability amplitude distributions at various time steps. Additionally, we implemented  a supervised neural network model to evaluate the effectiveness of data-driven techniques. The model learns to predict the squared modulus of the wavefunction \(|\psi(x, t)|^2\) given spatial and temporal coordinates. A comparative analysis demonstrates that the ML model can reproduce the qualitative structure and temporal progression of the quantum system with high accuracy. This study provides the synergy between quantum walk-based  simulation and machine learning for solving quantum dynamical equations.}

\keywords{Continuous-time quantum walk, Schr\"odinger equation, quantum simulation, machine learning, wavefunction dynamics.}

%%\pacs[JEL Classification]{D8, H51}

%%\pacs[MSC Classification]{35A01, 65L10, 65L12, 65L20, 65L70}

\maketitle

\section{Introduction}

Solving partial differential equations  is one of crucial study to understand physical systems in the theory of quantum mechanics and conputational fluid dynamics. The   Schr\"odinger equation plays a foundational role in soul of quantum mechanics. Traditional numerical methods such as finite difference, Crank-Nicolson, and spectral methods have been extensively studied to approximate its solution~\cite{griffiths2018introduction}, \cite{teschl2014quantum}. Recent advancements in quantum information techniques have opened up alternative formulations and simulation methods that mimic quantum evolution itself.

Quantum walks which is a quantum version of randon walks  provide a natural mathematical framework for simulating quantum dynamics on graphs~\cite{farhi1998quantum, childs2004spatial}. In a continuous time quantum walks (CTQW), the evolution of the quantum state is driven by a Hamiltonian operator computed from the graph, leading to coherent propagation similar to solutions of the Schr\"odinger equation. The resemblance between the CTQW evolution equation and the Schr\"odinger equation suggests that CTQWs can be used as a numerical solver for certain quantum dynamical systems. 

We utilized evolution process of continuous time quantum walks  to numerically solve the Schr\"odinger equation with a 1D harmonic potential. We approximate the spatial domain using a finite set of points. We formulate the Hamiltonian by combining the kinetic and potential energy terms, and propagate the initial wavefunction through the continuous time quantum walk evolution operator \( U(t) = \exp(-iHt) \). The findings exhibit that CTQW evolution effectively models quantum dynamics, such as oscillation and revival of wavepackets.

\subsection{Quantum Walks} 
Quantum walks  provide a powerful mathematical framework due to their quantum rooted properties. The evolution of a quantum walker   over a network containing nodes and their connectivity is based on quantum superposition, entanglement and interference. The two main types of quantum walks are the discrete-time quantum walks, which is product of two unitary operators: a coin and a shift operator, and continuous-time quantum walks, which evolve  under a time independent Hamiltonian. \\
 \indent In this research work, we attempt to simulate the time evolution of a particle's probability distribution using continuous time quantum walks, which offer to observe dynamic behavior over spatial domains with high resolution. These simulations serve as a baseline for comparing the accuracy and generalization capability of neural network-based approximations.
  The CTQW evolution operator
over the time parameter $t$, \( U(t) \), defined as
\begin{equation*} \label{eq1}
U(t) = e^{-iHt},
\end{equation*}
here $H$ represents Hamiltonian, $t$ time-parameter and $i$ complex unity.
\section{Mathematical Framework}

We solve the Schr\"odinger equation using a Continuous-Time Quantum Walk approach on a discretized 1D spatial grid. The governing equation is:
\[
i \frac{\partial \psi(x,t)}{\partial t} = \hat{H} \psi(x,t)
\]
where the Hamiltonian \( \hat{H} \) includes a kinetic and potential energy term:
\[
\hat{H} = -\frac{1}{2} \frac{\partial^2}{\partial x^2} + V(x)
\]
with the harmonic potential defined as:
\[
V(x) = \frac{1}{2} x^2
\]

\subsection*{Spatial Discretization}
Let the spatial domain be uniformly discretized as:
\[
x_i = x_0 + i \cdot \Delta x \quad \text{for } i = 0, 1, \dots, N-1
\]
where \( \Delta x = \frac{b - a}{N-1} \).

\subsection*{Laplacian Operator}
The second derivative is approximated using the finite difference scheme:
\[
\left. \frac{\partial^2 \psi}{\partial x^2} \right|_{x_i} \approx \frac{\psi_{i+1} - 2\psi_i + \psi_{i-1}}{(\Delta x)^2}
\]
Thus, the Laplacian operator \( {L} \in \mathbb{R}^{N \times N} \) is a tridiagonal matrix:
\[
{L} = \frac{1}{(\Delta x)^2} \left(
\begin{bmatrix}
-2 & 1 & 0 & \cdots & 0 \\
1 & -2 & 1 & \cdots & 0 \\
0 & 1 & -2 & \cdots & 0 \\
\vdots & \vdots & \vdots & \ddots & 1 \\
0 & 0 & 0 & 1 & -2
\end{bmatrix}
\right)
\]

\subsection*{Hamiltonian Matrix}
The discretized Hamiltonian is given by:
\[
{H} = -\frac{1}{2} {L} + {V}
\]
where \( {V} \) is a diagonal matrix with entries \( V(x_i) = \frac{1}{2} x_i^2 \).

\subsection*{Initial Wavefunction}
We initialize the wavefunction on the discrete spatial grid as a Gaussian,
\[
\psi_i(0) \;=\; \frac{e^{-x_i^2}}{\sqrt{\sum_j e^{-2x_j^2}\,\Delta x}},
\]
so that the discrete normalization condition
\(\sum_i |\psi_i(0)|^2 \,\Delta x = 1\) is satisfied.  
Here \(\Delta x = (b-a)/(N-1)\) is the grid spacing on the domain \([a,b]\), 
with Dirichlet boundary conditions \(\psi(a,t) = \psi(b,t) = 0\).  
We work in natural units \(\hbar = m = 1\).

\subsection*{CTQW Evolution}
The wavefunction evolves under the unitary operator
\[
\psi(t+\Delta t) \;=\; U(\Delta t)\,\psi(t), 
\qquad U(\Delta t) = e^{-i H \Delta t},
\]
where \(H\) is the discretized Hamiltonian matrix.  
To preserve unitarity in practice, we compute \(U(\Delta t)\) by 
diagonalizing \(H = Q \Lambda Q^{\top}\) and applying
\[
U(\Delta t) = Q \, e^{-i \Lambda \Delta t} Q^{\top}.
\]
This update is repeated for \(T\) timesteps:
\[
\psi^{(k+1)} = U(\Delta t)\,\psi^{(k)}, 
\qquad k = 0,1,\dots,T-1.
\]
Probability is conserved to numerical precision at all timesteps,
\(\sum_i |\psi_i(t)|^2 \Delta x = 1\).

\subsection*{Observation}
At each time step, we compute the probability density:
\[
|\psi(x,t)|^2 = \psi^*(x,t)\psi(x,t)
\]
to visualize the evolution of the wavefunction in space and time.

\begin{figure}[h]
    \centering
    \includegraphics[width=1\textwidth]{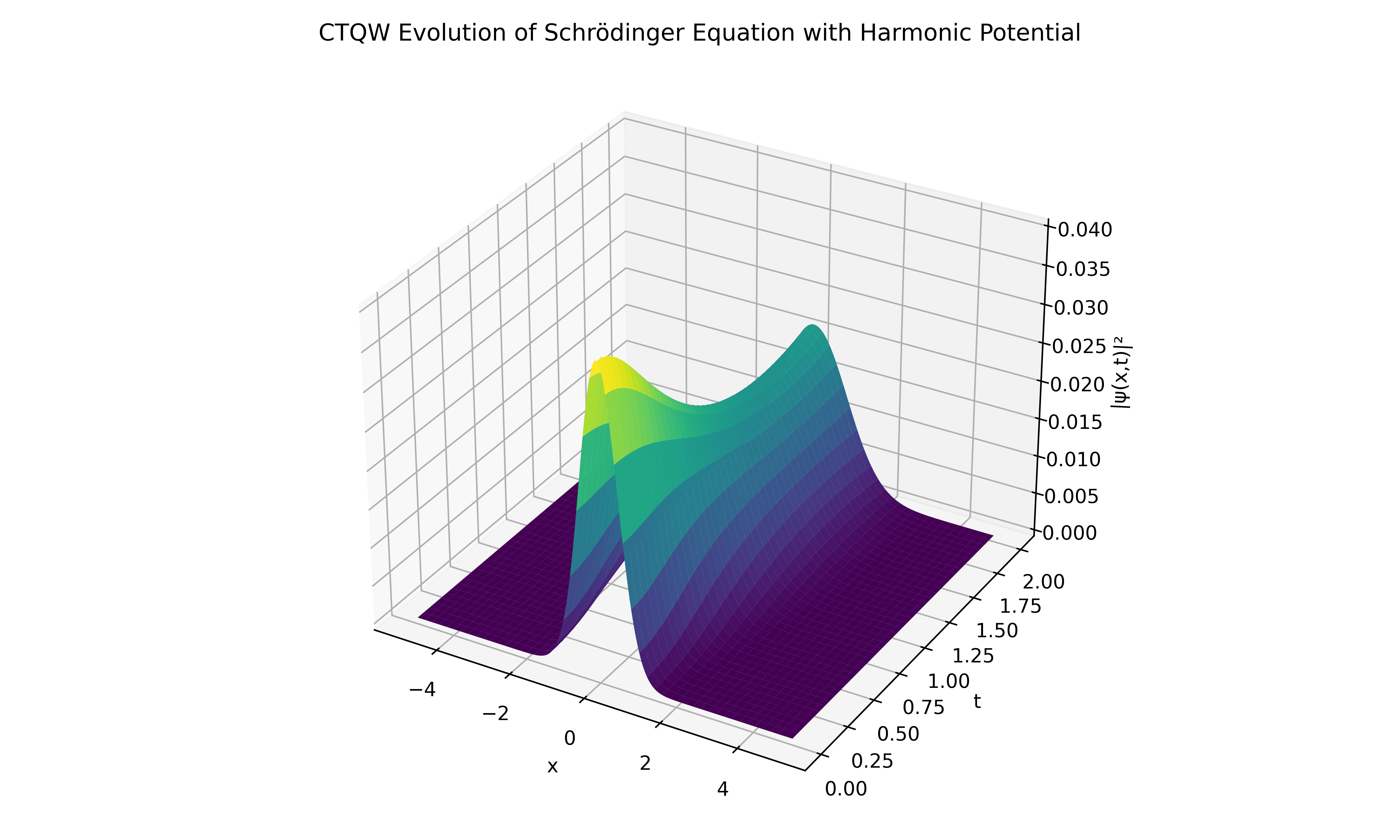}
    \caption{Evolution of the wavefunction \(|\psi(x, t)|^2\) over time for the Schr\"odinger equation with harmonic potential using CTQW.}
    \label{fig:ctqw}
\end{figure}
\begin{figure}[h]
    \centering
    \includegraphics[width=1\textwidth]{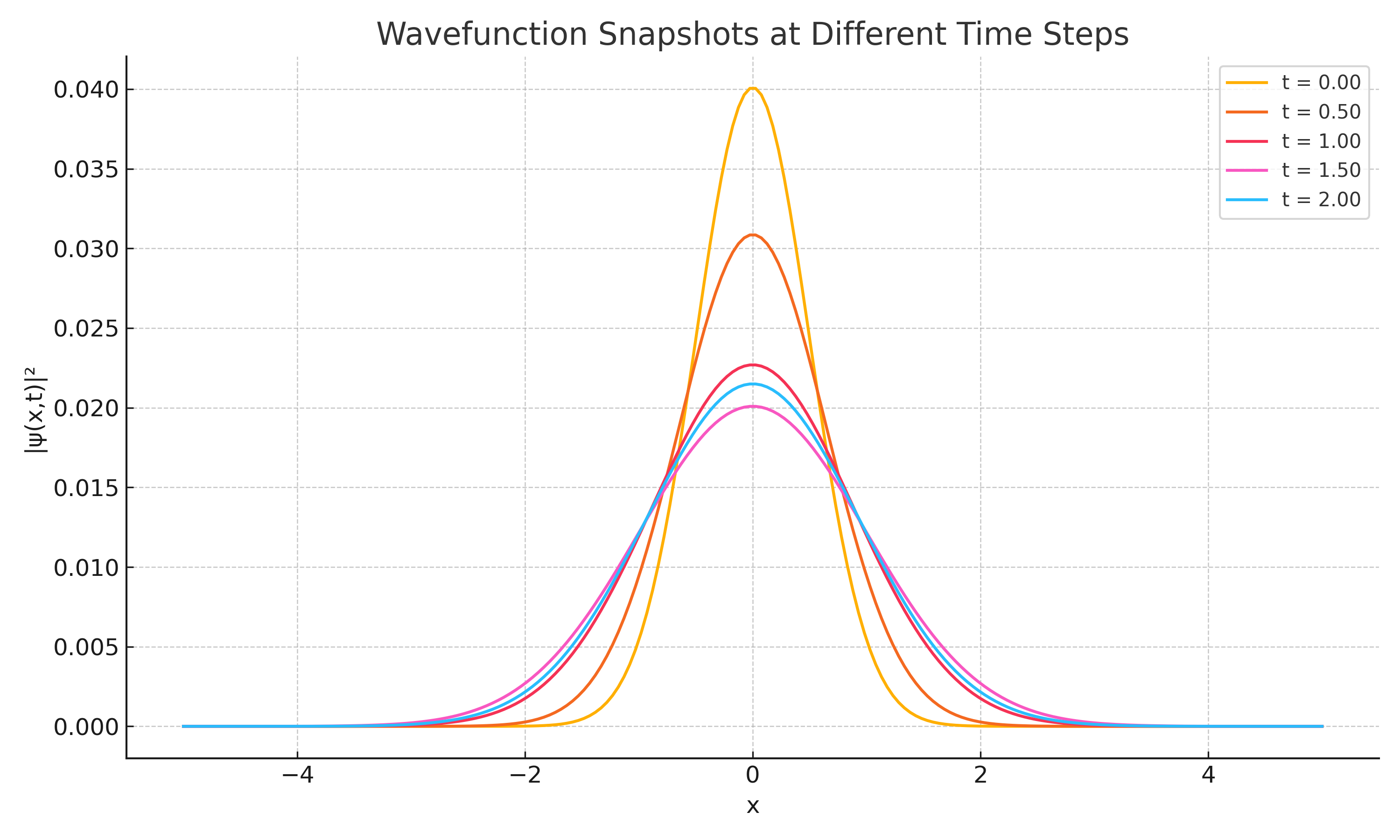}
    \caption{Evolution of the wavefunction \(|\psi(x, t)|^2\) over time for the Schr\"odinger equation with harmonic potential using CTQW.}
    \label{fig:ctqw}
\end{figure} 
\begin{table}[ht]
\centering
\caption{Wavefunction probabilities $|\psi(x,t)|^2$ at selected positions and time steps.}
\label{tab:wavefunction-values}
\begin{tabular}{c|c|c|c|c}
\toprule
Index & $x$ & $t=0.00$ & $t=0.50$ & $t=1.00$   \\
\midrule
0 & -5.000 & 7.73e-24 & 1.27e-15 & 2.66e-10   \\
1 & -4.950 & 2.10e-23 & 5.33e-15 & 1.08e-09   \\
2 & -4.899 & 5.66e-23 & 1.30e-14 & 2.48e-09  \\
3 & -4.849 & 1.51e-22 & 2.62e-14 & 4.54e-09  \\
4 & -4.799 & 3.97e-22 & 4.83e-14 & 7.36e-09  \\
\bottomrule
\end{tabular}
\end{table}
Values in Table \ref{tab:wavefunction-values} correspond to probability density samples at 
discrete nodes (not integrated probabilities)
\section{ Wavefunction Evolution Using Neural Networks}

Next, we explored a data-driven approach using machine learning to model and predict the evolution of the quantum wavefunction. We employed a Long Short-Term Memory (LSTM) neural network trained on the CTQW-generated dataset, which captures $|\psi(x,t)|^2$ values across discrete time slices.

\subsection{Dataset Preparation}

The dataset for training was constructed from the CTQW simulation output. We selected the probability distribution snapshots $|\psi(x,t)|^2$ for various time steps and arranged them such that each input sequence comprised a fixed number of past time steps (look-back window) and the corresponding output was the next time step's distribution. Data was normalized using min-max scaling to ensure effective learning.

\subsection{Neural Network Model}

We utilized a sequential LSTM model due to its ability to capture temporal dependencies in time-series data. The model architecture included:
\begin{itemize}
    \item One LSTM layer with 64 hidden units.
    \item One dense output layer with the same dimensionality as the spatial grid points.
    \item Mean Squared Error (MSE) as the loss function.
    \item Adam optimizer for training.
\end{itemize}

The model was trained for 100 epochs using a batch size of 1. The loss curve showed smooth convergence, indicating stable training dynamics.

\subsection{ Comparison}

\begin{figure}[H]
    \centering
    \includegraphics[width=1\textwidth]{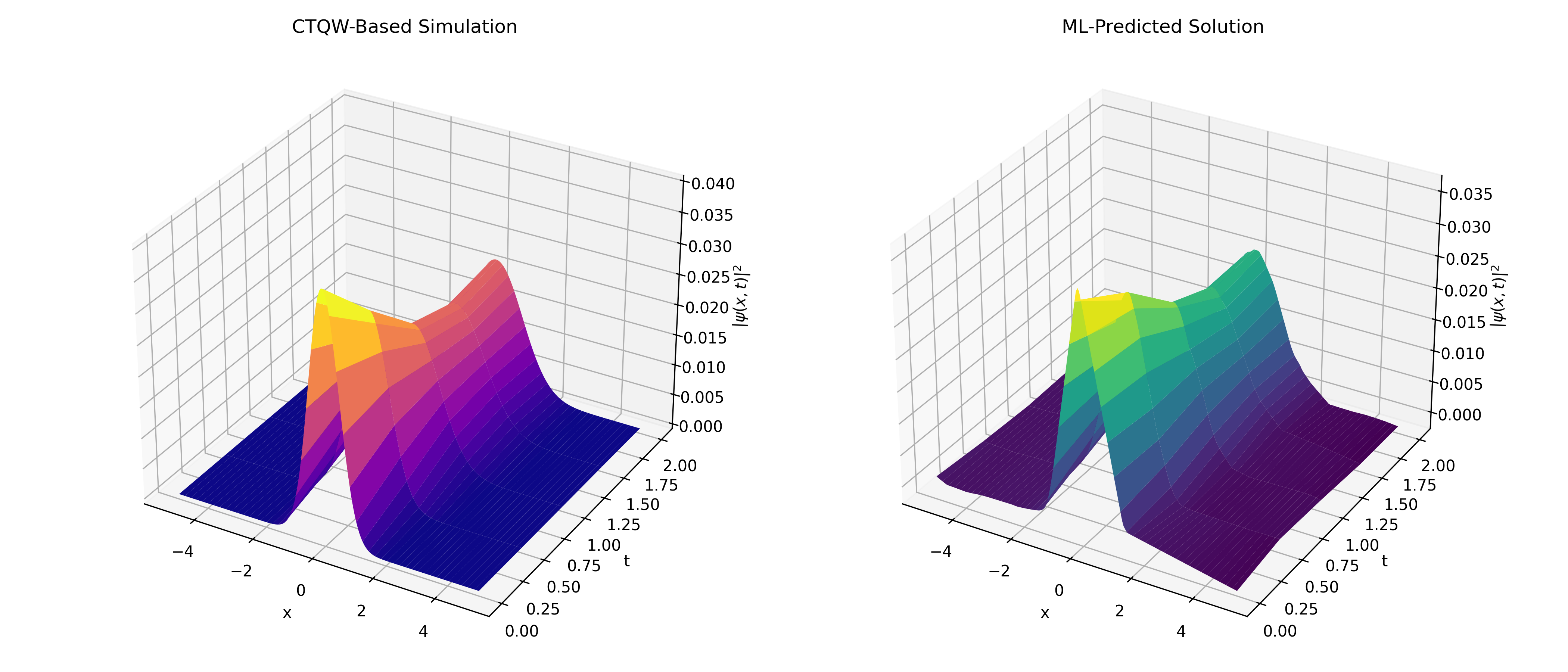}
    \caption{Comparison of wavefunction snapshots between CTQW simulation and ML prediction at different time steps.}
    \label{fig:ctqw_vs_ml}
\end{figure}

The trained model was evaluated on a test set, and its predictions were compared with CTQW-simulated wavefunctions. The results show good agreement in overall amplitude distribution and time evolution trends. Figure~\ref{fig:ctqw_vs_ml} compares selected time slices from both methods.

\section{Conclusion}

In this work, we explored the time evolution of the wavefunction governed by the Schr\"odinger equation with a harmonic potential using a Continuous-Time Quantum Walk (CTQW)-based numerical approach. We further utilized neural network model to learn and predict future wavefunction distributions from the simulated CTQW data. The predicted results captured the overall dynamics effectively, indicating the potential of machine learning as a surrogate model to forecast quantum evolution.  Future work may extend this framework to more complex potentials and higher-dimensional systems, enabling faster approximations and real-time analysis in quantum dynamics.

\nocite{*} 
\bibliography{reference}

\end{document}